\documentclass[a4paper]{jpconf}

\renewcommand{\slash}[1]{\displaystyle{\not} \hskip+0pt #1}

\usepackage{url}
\usepackage{multirow}
\usepackage{amsmath}
\usepackage{amssymb}

\usepackage{graphicx}
\begin{document}

\vspace*{-2cm}
\begin{flushright}
 LPT-Orsay 13-21 \\
 CFTP/13-005 \\
 PCCF-RI-13-03
\end{flushright}

\title{Potential of a linear collider for lepton flavour violation studies in the susy seesaw}

\author{Ant\'onio J.\ R.\ Figueiredo$^1$, A.\ Abada$^2$, J.\ C.\ Rom\~ao$^1$ and A.\ M.\ Teixeira$^3$}

\address{$^1$ CFTP, Departamento de F\'isica, Instituto Superior T\'ecnico,\\ Avenida Rovisco Pais, 1, 1049-001 Lisboa, Portugal}
\address{$^2$ Laboratoire de Physique Th\'eorique, CNRS -- UMR 8627,\\ Universit\'e de Paris-Sud 11, F-91405 Orsay Cedex, France}
\address{$^3$ Laboratoire de Physique Corpusculaire, CNRS/IN2P3 -- UMR 6533,\\ Campus des C\'ezeaux, 24 Av. des Landais, F-63177 Aubi\`ere Cedex, France}

\ead{ajrf@cftp.ist.utl.pt}

\begin{abstract}
	We study the potential of an $e^{\pm} e^-$ Linear Collider for charged lepton flavour violation studies 
	in a supersymmetric framework where neutrino masses and mixings are explained by a type-I seesaw. 
 	Focusing on $e$-$\mu$ flavour transitions, we evaluate the background from standard model and supersymmetric 
	charged currents to the $e \mu + \slash E_T$ signal. We study the energy dependence of both signal 
	and background, and the effect of beam polarisation in increasing the signal over background significance. 
	Finally, we consider the $\mu^- \mu^- + \slash E_T$ final state in $e^- e^-$ collisions that, despite being 
	signal suppressed by requiring two $e$-$\mu$ flavour transitions, is found to be a clear signature of 
	charged lepton flavour violation due to a very reduced standard model background. This contribution 
	summarises part of the work done in \cite{Abada:2012re}. 
\end{abstract}

\section{Introduction}

	A high-energy lepton collider (LC) \cite{Heuer:2012gi} offers the possibility of direct 
	production of lepton number ($L$) carrying heavy states (such as sleptons $\tilde l$ in supersymmetry) and of doing 
	precision physics profiting from low QCD activity. The possibility of polarising the colliding beams with 
	great accuracy makes the physics potential of a lepton collider even more ambitious. Indeed, by polarising 
	the initial beams, one is able to (approximately) project the chirality of the couplings intervening in the 
	primary production process, thus probing the chirality structure of an underlying new physics model. Polarisation can also be 
	used to suppress the background from Standard Model (SM) processes (e.g.\ from weak charged currents) to charged 
	lepton flavour violation (cLFV) signals. Moreover, threshold scans of chirality projected primary produced particles can be used to determine 
	their masses \cite{Feng:1998ud}. 

	Supersymmetry (SUSY) remains one of the most attractive extensions of the SM. Thus, a very appealing mechanism to explain 
	the smallness of neutrino masses and generate neutrino mixing is to consider a supersymmetric 
	seesaw. A type-I seesaw mechanism in which the right-handed (RH) neutrinos have masses close to the grand 
	unification (GUT) scale, generates neutrino masses and mixings with naturally large Yukawa couplings, 
	and can offer an explanation for the observed baryon asymmetry of the universe. However, the drawback 
	of such a set-up is that it is very hard to probe since the very heavy RH neutrinos cannot be produced at 
	colliders. On the other hand, in a supersymmetric type-I seesaw (usually dubbed SUSY seesaw) the radiative 
	corrections generate flavour violation in the slepton sector \cite{Borzumati:1986qx}, giving 
	additional sources of LFV at low energy. At a high-energy LC, sleptons can be copiously produced and their decays 
	lead to potentially observable cLFV final states, thus providing a unique probe of this mass 
	generation mechanism. 

	SUSY-seesaw induced cLFV at a LC has been studied in \cite{Deppisch:2003wt,Deppisch:2004pc} focusing on 
	$\tau$-$\mu$ flavour violation, while in \cite{Carquin:2011rg} slepton driven cLFV was also considered but 
	without relying on a particular origin of slepton mixing.  

	Motivated by the excellent muon reconstruction capabilities and the recently improved upper-limit on 
	$\mu\to e\gamma$ \cite{Adam:2011ch}, we studied the SUSY seesaw induced $\mu$-$e$ 
	cLFV and its discovery potential at a LC working with polarisable beams \cite{Abada:2012re}.

\section{The SUSY seesaw}

	The SUSY seesaw model comprises the minimal supersymmetric standard model (MSSM) 
	extended by three\footnote{We assume that the number of RH neutrino generations mimics the number of generations of SM fermions. In fact, 
	two are sufficient in order to comply with neutrino oscillation parameters.} chiral superfields 
	$\hat N^c_i \sim \left( \nu^c, \tilde \nu^\dagger_R \right)_i$, so-called RH neutrino superfields, that are singlets under the SM gauge group.
	The leptonic part of the superpotential is given by 
	\begin{equation}
		\mathcal{W}^{\text{lepton}} = \frac{1}{2} \hat N^c M_N \hat N^c + \hat N^c Y^\nu \hat L \hat H_2 + \hat E^c Y^l \hat L \hat H_1 \,.
	\end{equation}
	Hereafter we work in a flavour basis where the charged lepton Yukawa couplings $Y^l$ and the RH neutrino mass matrix $M_N$ 
	are diagonal. The slepton soft breaking terms are 
	\begin{eqnarray}
		\mathcal{V}^{\text{slepton}}_{\text{soft}} & = & m^2_{\tilde L} \tilde \ell_L \tilde \ell^*_L + m^2_{\tilde E} \tilde \ell_R \tilde \ell^*_R + m^2_{\tilde \nu_R} \tilde \nu_R \tilde \nu^*_R \nonumber\\
		&& + \left( A_l H_1 \tilde l_L \tilde l^*_R + A_\nu H_2 \tilde \nu_L \tilde \nu^*_R + B_\nu \tilde \nu_R \tilde \nu_R + \text{H.c.} \right) \, .
	\end{eqnarray}
	Our analysis is conducted in a framework where SUSY breaking is flavour blind (as in minimal supergravity mediated SUSY breaking) and 
	the soft breaking parameters satisfy certain universality conditions at a high-energy scale, which we take to be the gauge coupling 
	unification scale $M_{\text{GUT}} \sim 10^{16}$ GeV: 
	\begin{eqnarray}
		&& \left( m_{\tilde L} \right)^2_{ij} = \left( m_{\tilde E} \right)^2_{ij} = \left( m_{\tilde \nu_R} \right)^2_{ij} = m^2_0 \delta_{ij} \,,\\
		&& \left( A_l \right)_{ij} = A_0 \left( Y^l \right)_{ij} \,, \left( A_\nu \right)_{ij} = A_0 \left( Y^\nu \right)_{ij} \,,
	\end{eqnarray}
	where $m_0$ and $A_0$ are the universal scalar soft-breaking mass and trilinear couplings of the constrained MSSM (cMSSM), and $i,j$ 
	denote lepton flavour indices ($i,j = 1,2,3$). 

	After electroweak (EW) symmetry breaking, the neutrino mass matrix is approximately given by $m_\nu \simeq -{m^\nu_D}^T M^{-1}_N m^\nu_D$, 
	where $m^\nu_D = Y^\nu v_2$ and $v_i$ is the vacuum expectation value of $H_i$ ($v_{1(2)} = v \cos(\sin)\beta$, with $v = 174$ GeV). 
	A convenient means of parametrizing the neutrino Yukawa couplings, while at the same time allowing to accommodate the experimental data, 
	is given by the Casas-Ibarra parametrization \cite{Casas:2001sr}, which reads at the seesaw scale $M_N$ 
	\begin{equation}
		Y^\nu v_2 = m^\nu_D = i \sqrt{M^{\text{diag}}_N} R \sqrt{m^{\text{diag}}_\nu} U^\dagger_{\text{MNS}} \,, \label{eq.CasasIbarra}
	\end{equation}
	where $U_{\text{MNS}}$ is the light neutrino mixing matrix and $R$ is a $3 \times 3$ complex orthogonal matrix that encodes the possible 
	mixings involving RH neutrinos. We use the standard parametrization for the $U_{\text{MNS}}$, with the three mixing angles in the intervals 
	favoured by current data \cite{GonzalezGarcia:2012sz}.

	\subsection{Flavour violation in the slepton sector} 

		Due to the non-trivial flavour structure of $Y^\nu$, the running from $M_{\text{GUT}}$ down to the seesaw scale will induce 
		flavour mixing \cite{Borzumati:1986qx} in the otherwise approximately flavour conserving slepton soft breaking terms. 
		This running is more 
		pronounced in the soft breaking terms involving slepton doublets since these have local interactions with the RH (s)neutrinos. 
		At leading order, the flavour mixing induced by these radiative corrections has the form 
		\begin{eqnarray}
			&& \left( \Delta m^2_{\tilde L} \right)_{ij} = -\frac{1}{8\pi^2} \left( 3 m^2_0 + A^2_0 \right) \left( {Y^\nu}^\dagger L Y^\nu \right)_{ij} \,,\\
			&& \left( \Delta A_l \right)_{ij} = -\frac{3}{16\pi^2} A_0 Y^l_{ij} \left( {Y^\nu}^\dagger L Y^\nu \right)_{ij} \,;\, L_{kl} \equiv \log\left(\frac{M_{\text{GUT}}}{M_{N_k}}\right) \delta_{kl} \,.
		\end{eqnarray}
		The amount of flavour violation in the slepton sector is encoded in $\left( {Y^\nu}^\dagger L Y^\nu \right)_{ij}$ which, 
		as made explicit by equation \eqref{eq.CasasIbarra}, is related to high- and low-energy neutrino parameters.

\section{cLFV in $e^{\pm} e^-$ collisions}

	At collider energies, i.e.\ at energies of the order of the TeV, the (high-scale) SUSY seesaw with the aforementioned 
	assumptions for SUSY breaking can be interpreted as a minimal deviation from the cMSSM that allows for flavour mixing in 
	the slepton soft breaking parameters, since processes mediated by RH (s)neutrinos are greatly suppressed due to the 
	very high seesaw scale. We stress that, contrary to other analysis, the origin of this flavour mixing is rooted on, albeit 
	not singly determined by, neutrino oscillations.

	In the SUSY seesaw model, as in the MSSM, the sparticle production in $e^+ e^-$ collisions are dominated by the following 2-body processes 
	\begin{equation}
		e^+ e^- \to \tilde \ell^+_i \tilde \ell^-_j\,,~ \chi^0_A \chi^0_B\,,~ \chi^+_A \chi^-_B\,,~ \tilde\nu^*_i \tilde\nu_j \,. \label{eq.EpEm2body}
	\end{equation}
	For $e^- e^-$ collisions, the available channels are considerably restricted, and we have 
	\begin{equation}
		e^- e^- \to \tilde \ell^-_i \tilde \ell^-_j \,. \label{eq.EmEm2body}
	\end{equation}

	Since SUSY-seesaw induced cLFV final states are dominated by slepton decays via slepton-lepton-(EW gaugino) interactions, it is useful to 
	distinguish two cases with respect to the slepton/EW gaugino mass hierarchy. A dark-matter motivated scenario in which 
	sleptons are heavier than the EW gauginos occurs in the so-called ``Higgs funnel'' region, while the opposite hierarchy 
	happens in the so-called co-annihilation region. We thus define two types of points to guide the subsequent analysis 
	\begin{itemize}
		\item F points: $m_{\tilde \ell, \tilde \nu} > m_{\chi^0_2, \chi^+_1}$ and $m_{\chi^0_2, \chi^+_1} > m_{\tilde\tau_1}$; 
		\item C points: $m_{\tilde \ell, \tilde \nu} < m_{\chi^0_2, \chi^+_1}$ and $m_{\chi^0_1} \lesssim m_{\tilde\tau_1}$.
	\end{itemize}
	In table \ref{tb.points} we give two C- and one F-type points that will be used in the numerical analysis. 

			\begin{table}[h]
			\caption{\label{tb.points} Representative points used in the numerical analysis.} 
				\begin{center}
				\lineup
				\begin{tabular}{*{5}{l}}
				\br                              
					& C$_1$ & C$_2$ & F \cr 
				\mr
					$m_0$ (GeV) 	& 150 & 200 & 750 \cr
					$M_{1/2}$ (GeV)	& 727.9 & 949.2 & 872.1 \cr 
					$\tan\beta$	& \010 & \010 & \052 \cr
					$A_0$ (GeV)	& \0\00 & \0\00 & \0\00 \cr 
					sign($\mu$)	& \0\01 & \0\01 & \0\01 \cr
				\br
				\end{tabular}
				\end{center}
			\end{table}

	In the model under consideration, R-parity is conserved implying that the final states we are interested in, i.e.\ 
	those with intervening sleptons, have an even number of $\chi^0_1$ (the lightest supersymmetric particle, the LSP) which 
	escape the detector without interacting. 
	Hence, we focus our attention on cLFV in processes with di-lepton final states plus missing transverse momenta,  
	$\slash E_T$. The main source of background arises from weak charged currents producing neutrinos which, analogously 
	to the LSP, escape undetected. These are of two types: the SM charged current backgrounds (type B) in which all $\slash E_T$ 
	is due to neutrinos; SUSY charged current backgrounds (type C) in which $\slash E_T$ contains both neutrinos and LSPs. 
	Here, we study the following possibilities: 
	\begin{eqnarray}
		e^\pm e^- \to \left\{ \begin{array}{ll}
				e^\pm \mu^- + 2 \chi^0_1 & \text{(A)} \\
				e^\pm \mu^- + 2 \chi^0_1 + (2,4)\nu & \text{(B)} \\
				e^\pm \mu^- + (2,4)\nu & \text{(C)}
			\end{array} \right.
	\end{eqnarray}
	where (A) is the signal. In table \ref{tb.bckgs} we classify the main sources of (B) and (C) backgrounds in $e^+ e^-$ 
	collisions, in which we allow for leptonically decaying $\tau$s, $W$s and $Z$s. Backgrounds in $e^- e^-$ collisions can 
	be similarly classified. However, $e^- e^- \to W^- W^-$ is very small, being suppressed by powers of $m_\nu / M_W$, and 
	$e^- e^- \to \tau^-\tau^-$ vanishes.

	\begin{table}[h]
		\caption{\label{tb.bckgs} SUSY charged current backgrounds (B) and SM charged current backgrounds (C), with 
					  the corresponding total cross section (order of magnitude) for unpolarised beams, for 
					  the signal $e^+ \mu^- + 2\chi^0_1$.} 
		\begin{center}
		\lineup
			\begin{tabular}{lll}
				\br
				\multicolumn{1}{l}{\multirow{3}{*}{\begin{tabular}{c}SUSY bckg (type B)\\ $\lesssim 5$ fb\end{tabular}}} & $0\tau$ 	   & $e^+ \mu^- + (\bar\nu\nu,2\bar\nu\nu) + 2\chi^0_1$	\\ \cline{2-3}
				\multicolumn{1}{l}{} 		   		 & $\tau$+$0\nu$ & $({e^+ \tau^-}, {\tau^+ \mu^-}, \tau^+\tau^-) + 2\chi^0_1$ \\ \cline{2-3}
				\multicolumn{1}{l}{}		   		 & $\tau$+$2\nu$ & $(e^+ \tau^-, \tau^+ \mu^-) + \bar\nu\nu + 2\chi^0_1$				\\ \mr
				\multicolumn{1}{l}{\multirow{3}{*}{\begin{tabular}{c}SM bckg (type C)\\ $\approx 10^2$ fb\end{tabular}}} & $W$-strahlung & $W^- (e^+,\tau^+) \nu$, $W^+ (\mu^-,\tau^-) \bar\nu$ \\ \cline{2-3}
				\multicolumn{1}{l}{}			  & $W$-pair	 & $W^+ W^-$						\\ \cline{2-3}
				\multicolumn{1}{l}{}			  & {$\tau$}-pair	 & $\tau^+ \tau^-$				\\ \br
			\end{tabular}
		\end{center}
	\end{table}

	SM charged current backgrounds could {\it in principle} be reduced by devising kinematical cuts that rely 
	on the fact that neutrinos are much lighter than the LSP and that the seeding processes have different topologies. 
	However, that analysis is beyond the scope of our work. In our work \cite{Abada:2012re} we conduct a phenomenological analysis, focusing on theoretical 
	estimations of the expected number of events at a LC operating at a given centre of mass energy with the possibility of polarised beams.

	More precisely, our analysis is based on an algorithmic calculation of the possible production and decay modes, considering that the majority 
	of the events proceeds from an on-shell primary production (so that there are no interference effects between the different 
	contributions), with subsequent two-body cascade decays (the exception being 3-body decays of the $\tau$). 

	For sparticle primary production, we have considered the aforementioned channels\footnote{Total lepton number violating processes 
	such as $e^+ e^- \to \tilde \nu \tilde \nu$ and 
	$e^- e^- \to \chi^-_A \chi^-_B$ are severely suppressed by the seesaw scale.}, i.e.\ those listed in equations \eqref{eq.EpEm2body} and 
	\eqref{eq.EmEm2body}, while for SM primary production we have 
	considered those listed in table \ref{tb.bckgs}, i.e.\ $W$-strahlung for both $e^+ e^-$ and $e^- e^-$ collisions, and 
	$W$- and $\tau$-pair productions for $e^+ e^-$ collisions. 

\begin{figure}[h]
			\begin{minipage}{150mm}
				\includegraphics[width=75mm]{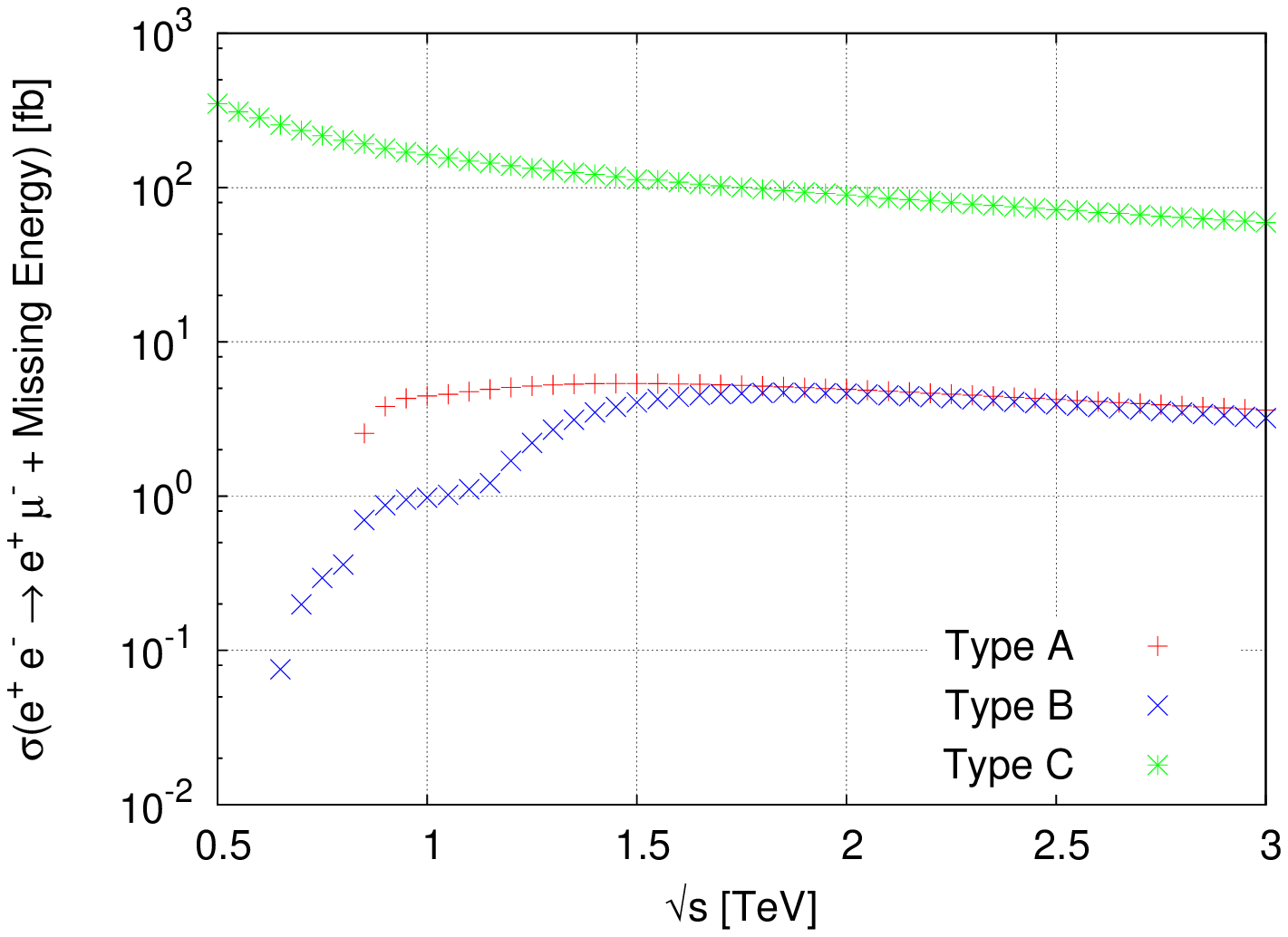}
				\hspace{2pc}\includegraphics[width=75mm]{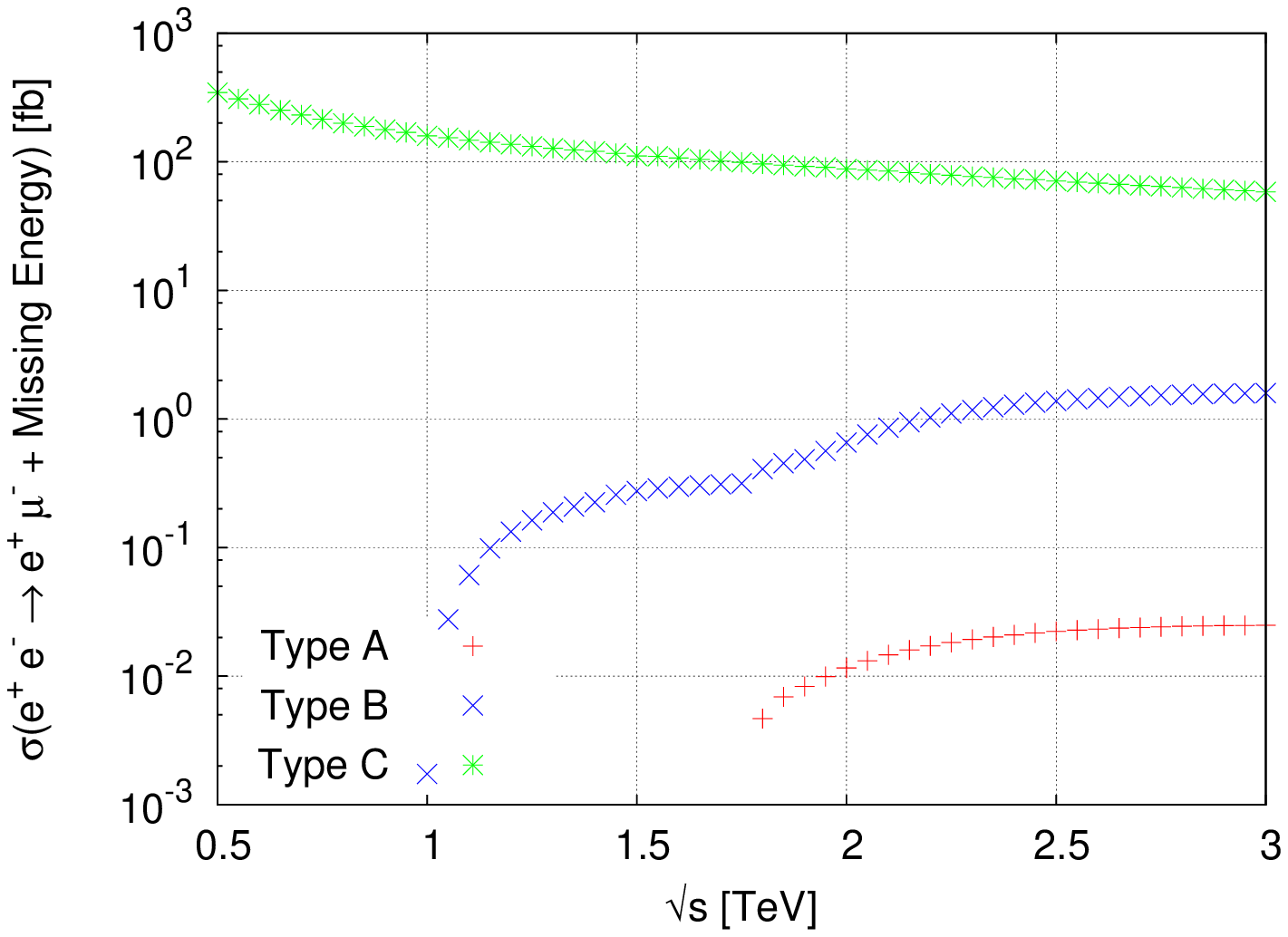}
				\caption{\label{fig.C1Fsqrts} Cross sections for $e^+ e^- \to e^+ \mu^- + \slash E_T$ 
							(with $\slash E_T = 2 \chi^0_1, 2\chi^0_1 + (2,4)\nu, (2,4)\nu$), 
							for points C$_1$ and F (left and right panels, respectively) as 
							a function of the centre of mass energy $\sqrt s$. 
							We fix $M_R = 10^{12}$ GeV, and denote the signal (A) with red crosses, 
							the SUSY charged current background (B) with blue times, and the SM 
							charged current background (C) by green asterisks. We have taken a 
							degenerate right-handed neutrino spectrum and set $\theta_{13} = 10^\circ$.} 
							
			\end{minipage}
		\end{figure}

	In figure \ref{fig.C1Fsqrts} we show the results for $e^+ e^-$ unpolarised beams for 
	point C$_1$ (left hand side) and F (right hand side). The cross sections for the signal (red crosses), SUSY background 
	(blue times) and SM background (green asterisks) are given as a function of the centre of mass energy, $\sqrt s$. We have 
	taken a degenerate RH neutrino spectrum with $M_R = 10^{12}$ GeV, and $\theta_{13} = 10^\circ$.

	As can be seen, the SM background, consisting mainly of $W$-strahlung and $W$-pair production, dominates the signal by approximately 
	one order of magnitude for C$_1$ and three orders of magnitude for F. Moreover, the SUSY background dominates (is comparable to) 
	the signal in F (C$_1$). To understand these two observations, notice that the principal decay modes of sleptons and EW gauginos 
	in F-type points are given by 
	\begin{eqnarray}
		&& (\tilde \ell^-,\tilde\nu^*) \to \chi^0_2 (\ell^-,\bar\nu)\,,~ \chi^-_1 (\nu, \ell^+)\,,~ (\ell^-,\bar\nu) \chi^0_1 \,, \\
		&& \chi^-_1 \to \tilde\tau^-_1 \bar\nu, W^- \chi^0_1 \,, \\
		&& \chi^0_2 \to \tilde\tau^-_1 \tau^+, Z \chi^0_1 \,.
	\end{eqnarray}
	while for C-type points we have 
	\begin{eqnarray}
		&& \chi^-_1 \to \tilde \ell^- \bar\nu\,,~ \tilde\nu^* l^- \,, \\
		&& \chi^0_2 \to \tilde \ell^\pm l^\mp\,,~ \tilde \nu \bar\nu\,,~ \tilde \nu^* \nu \,, \\
		&& (\tilde \ell^-,\tilde \nu) \to (\ell^-, \nu) \chi^0_1 \,.
	\end{eqnarray}
	Thus, solely from the width of the decays we expect $\Gamma_{\text{total}}(\tilde \ell)_F > \Gamma_{\text{total}}(\tilde \ell)_C$, 
	which, for the same amount of flavour violation, gives that the expected signal cross sections for F points should be lower than for C points. 
	In fact, we find that typically $\left( \text{BR($\tilde e_L \to e \chi^0_1$)}_C \right)^2 \approx 10 \times \left( \text{BR($\tilde e_L \to e \chi^0_1$)}_F \right)^2$.

	Additionally, we observe that in C-type points, by taking a centre of mass energy above the $\tilde \ell_L \tilde \ell_R$ production 
	threshold and below the production threshold of both $\chi^+_1 \chi^-_1$ and $\chi^0_2 \chi^0_2$, it is possible to evade the largest 
	contribution from SUSY charged current backgrounds. For C$_1$ this happens above $\sqrt s \approx 800$ GeV and below $\sqrt s \approx 1200$ GeV, 
	as can be inferred from figure \ref{fig.C1Fsqrts}.

	\subsection{Beam polarisation effect} 

		Since $W$-pair production and $W$-strahlung backgrounds dominate the signal in $e^+ e^-$ 
		collisions, it is interesting to consider how beam polarisation can be exploited to suppress their 
		contribution without compromising the signal. We will do this by considering limiting cases, that is, 
		$100$\% polarisations. 

		Since slepton flavour mixing occurs predominantly in the LL slepton sector, i.e.\ via 
		decays of mostly left-handed slepton mass eigenstates, to avoid suppressing the $e^+ \mu^-$ signal 
		we require the $e^-$ beam to be $100\%$ left-polarised, while no constraint is placed on the positron 
		beam. 

		Constraints on the positron beam arise by requiring the suppression of both $W$-pair production and 
		$W$-strahlung production cross sections. In $W$-pair production, the s-channel is strongly dominated by 
		vector boson interactions which can be suppressed by taking equally polarised beams. The t-channel 
		is suppressed either by $e^+_L$ or $e^-_R$ beams. Therefore, we can suppress both channels by left-polarising 
		the positron beam. Regarding $W$-strahlung, a maximal suppression would require $e^+_L e^-_R$ which would 
		strongly suppress the signal. Nevertheless, the choice $e^+_L e^-_L$ is preferable to $e^+_R e^-_L$ one. 

\begin{figure}[h]
			\begin{center}
				\includegraphics[width=75mm]{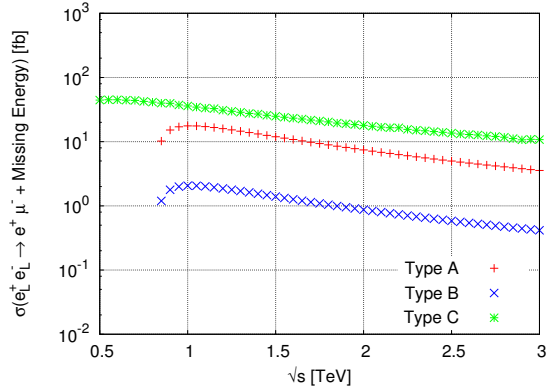}
			\end{center}
			\caption{\label{fig.C1LL} Cross section for $e^+ e^- \to e^+ \mu^- + \slash E_T$ (with $\slash E_T = 2\chi^0_1,2\chi^0_1+(2,4)\nu,
						(2,4)\nu$), for point C$_1$, as a function of the centre of mass energy $\sqrt s$, for $100$\% LL 
						polarised beams. We have taken degenerate right-handed neutrino spectrum with $M_R = 10^{12}$, and set 
						$\theta_{13} = 10^\circ$. Line and colour code as in figure \ref{fig.C1Fsqrts}.}
		\end{figure}

		In figure \ref{fig.C1LL} we show the result of left-polarising both positron and electron beams for 
		point C$_1$,  with parameters and line colour code as in figure \ref{fig.C1Fsqrts}. Clearly, 
		the signal is enhanced by a factor of $\approx 4$ near the ($\tilde e^+_R \tilde e^-_L$) production 
		threshold, while at higher energies the enhancement becomes smaller. The low energy tail of the SUSY background 
		``disappears'', as is understandable from the fact that such processes  proceed via $\tilde \tau^+_1 \tilde \tau^-_1$ 
		primary production. Moreover, both $\chi^+_1 \chi^-_1$ and $\chi^0_2 \chi^0_2$ primary production are 
		greatly suppressed. As expected, the SM background is also suppressed, becoming comparable to the signal. 

	\subsection{Probing the SUSY seesaw} 

		We now discuss how the observation of a signal would allow to probe the underlying mechanism 
		of slepton flavour mixing. If compatible 
		with seesaw predictions, an observation would indeed strengthen the seesaw hypothesis, since a single mechanism would be able to 
		explain many observables: the smallness of neutrino masses while generating neutrino mixings; and 
		collider cLFV. Moreover, low energy observables such as CR($\mu$-$e$, N) and 
		BR($\mu \to e \gamma$) would serve to further strengthen the hypothesis, provided that 
		the predictions would be compatible with observations/exclusion limits.

\begin{figure}[h]
			\begin{center}
				\includegraphics[width=75mm]{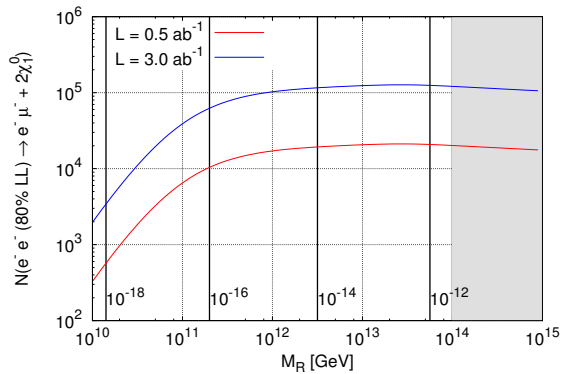}
			\end{center}
			\caption{\label{fig.nevents} Number of events for $e^- e^- \to e^- \mu^- + 2\chi^0_1$ for point 
						C$_2$ as a function of the seesaw scale $M_R$, for $(P_{e^-},P_{e^-})=(-80\%,-80\%)$ 
						polarised beams. In both cases, we fix $\sqrt s = 2$ TeV, and we have taken a degenerate 
					right-handed neutrino spectrum, with $\theta_{13}=10^\circ$. Vertical lines denote the $M_R$-corresponding 
					value of BR($\mu \to e \gamma$) while the (grey) shaded region represents values of $M_R$ 
					already excluded by the present experimental bound on BR($\mu \to e \gamma$).}
		\end{figure}

		In figure \ref{fig.nevents} we display the expected number of $e^- \mu^- + 2\chi^0_1$ events 
		from $80$\% LL polarised electron beams, as a function of the seesaw scale, $M_R$, for point 
		C$_2$. For illustrative purposes we have chosen $\sqrt s = 2$ TeV. 

		Due to the high number of expected events, if no $e^- \mu^-$ signal event is observed 
		at an LC, then a high-scale SUSY seesaw is clearly disfavoured. Supposing that events 
		are indeed observed, two possibilities arise. If the observed number of events is not accommodated 
		by predictions, then it is likely that either the SUSY seesaw is not the unique source of LFV or 
		that another mechanism for neutrino mass generation is at work. If the observed 
		number of events can be accommodated by predictions, as illustrated in figure \ref{fig.nevents} for 
		point C$_2$, then one can constrain (or even hint at) the seesaw scale. For example, if SUSY is realised 
		with a C$_2$-like spectrum and more than $10^5$ signal events are observed for a total integrated 
		luminosity of approximately $3$ ab$^{-1}$, then the seesaw scale, $M_R$, should be above $10^{12}$ GeV. 
		Moreover, evading the exclusion limits on $\mu\to e\gamma$ sets an upper bound on this scale, so that 
		we would be led to $10^{12}$ GeV $\lesssim$ $M_R$ $\lesssim$ $10^{14}$ GeV. If we now suppose that $\mu\to e\gamma$ 
		is observed at MEG, e.g.\ with a branching ratio of the order of $10^{-12}$, we would expect 
		a number of $e^-\mu^-$ signal events below $10^{5}$. This incompatibility could be due to (unaccounted for) 
		destructive interferences in collider processes, or additional sources of LFV that only contribute to low-energy 
		observables.

	\subsection{$\mu^-\mu^-$ golden channel}

		Same sign di-muon final states may possibly be a ``golden channel'' for the 
		detection of cLFV at an LC. From an experimental point of view, the efficiency of 
		the muon detectors can be fully explored when looking for di-muon 
		final states. On the theoretical side, higher signal significances are expected, 
		in particular due to very suppressed SM backgrounds. 

\begin{figure}[h]
			\begin{center}
				\includegraphics[width=75mm]{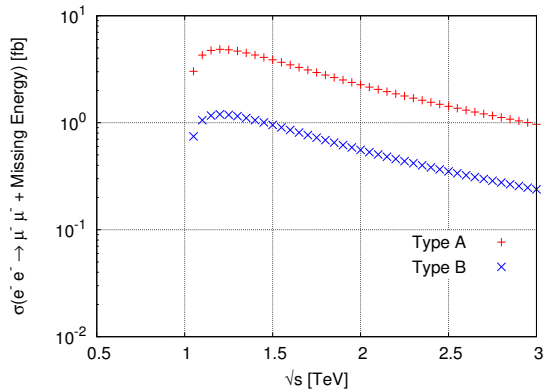}
			\end{center}
			\caption{\label{fig.mumu} Cross section for $e^- e^- \to \mu^- \mu^- + \slash E_T$ (with $\slash E_T = 2\chi^0_1,2\chi^0_1+(2,4)\nu$), 
						for point C$_1$ as a function of the centre of mass energy, $\sqrt s$, for the case of 
						unpolarised beams. The signal is denoted by red crosses and the SUSY background by 
						green asterisks. We have taken a degenerate right-handed neutrino spectrum with $M_R = 10^{12}$ GeV 
						and set $\theta_{13} = 10^\circ$.}
		\end{figure}

		In figure \ref{fig.mumu} we display the $e^- e^- \to \mu^-\mu^- + \slash E_T$ cross section  
		for the signal and the SUSY background, considering unpolarised beams, as a function of the 
		centre of mass energy. 

		The SM background, not displayed in figure \ref{fig.mumu}, is dominated by double-$W$-strahlung 
		production, i.e.\ $e^- e^- \to W^- W^- \nu \nu \to \mu^-\mu^- 2(\bar\nu\nu)$, whose cross section is 
		of the order of $1$ fb. Therefore, we confirm that the $\mu^-\mu^-$ signal event is indeed much cleaner than 
		$e^-\mu^-$ (and $e^+\mu^-$ in $e^+e^-$ collisions).

\section{Concluding remarks}

		A high-energy lepton collider offers an enormous potential for cLFV discovery. 
		Beam polarisation can be 
		instrumental in maximising the signal significance, rendering the signal ``visible'' 
		in a large part of the high-scale SUSY seesaw parameter space, even without dedicated 
		cuts, which could improve the observation prospects even further. We have also pointed out 
		that cLFV discovery at an LC, complemented with low-energy LFV observables, could substantiate 
		or disfavour the high-scale SUSY seesaw. 
		Finally, we commented on the truly remarkable channel for cLFV discovery: $e^- e^- \to \mu^- \mu^- + 2\chi^0_1$.

\ack

	This work has been done partly under the ANR project CPV-LFV-LHC {NT09-508531}. 
	The work of  A.\ J.\ R.\ F.\  has been supported by {\it Funda\c c\~ao para a Ci\^encia e a 
	Tecnologia} through the fellowship SFRH/BD/64666/2009. A.\ A.\, A.\ J.\ R.\ F.\ and A.\ M.\ T.\ 
	acknowledge partial support from the European Union FP7 ITN INVISIBLES (Marie Curie Actions, 
	PITN-GA-2011-289442). A.\ J.\ R.\ F.\ and J.\ C.\ R.\ also acknowledge the financial support from
	the EU Network grant UNILHC PITN-GA-2009-237920 and from {\it Funda\c{c}\~ao para a Ci\^encia 
	e a Tecnologia} project PEst-OE/FIS/UI0777/2011 and grants CERN/FP/116328/2010 and PTDC/FIS/102120/2008.

\end{document}